\documentclass[a4paper]{jpconf}
\usepackage{epsfig}
\begin{document}
\title{Jet quenching and heavy quarks}

\author{Thorsten Renk}

\address{Department of Physics, P.O. Box 35, FI-40014 University of Jyv\"askyl\"a, Finland}

\ead{thorsten.i.renk@jyu.fi}

\begin{abstract}
Jet quenching and more generally physics at high transverse momentum $P_T$ scales is a cornerstone of the heavy-ion physics program at both RHIC and LHC. In this work, the current understanding of jet quenching in terms of a QCD shower evolution being modified by the surrounding medium is reviewed along with the evidence for this picture from light parton high $P_T$ observables. Conceptually, the same QCD shower description should also be relevant for heavy quarks, but with several important modifications introduced by the quark masses. Thus especially in the limit of small jet energy over quark mass $E_{jet}/m_q$, the relevant physics may be rather different from light quark jets, and several attempts to explain the observed phenomenology of heavy quarks at high $P_T$ are discussed here.
\end{abstract}

\section{The QCD shower in vacuum}

Perturbative Quantum-Chromodynamics (pQCD) is used very successfully to understand particle production at high $P_T$ in proton-proton (p-p) collisions by factorizing the perturbatively calculable hard process from the non-perturbative initial and final state which is parametrized by universal objects, the parton distributions (PDFs) and fragmentation functions (FFs) respectively. It has been realized early on that hard processes can hence be utilized in the context of heavy-ion collisions by providing a probe with a perturbatively calculable rate, which then allows medium tomography via the parton final state interaction with the bulk QCD medium created in such collisions \cite{radiative1}.

The partons emerging from a hard process are  highly virtual and hence utilize the phase space provided by their virtuality for  intense QCD radiation, known as a parton shower, which after hadronization is experimentally observed as a hadron jet. Due to the hard virtuality scale, such showers are perturbatively tractable and simulated e.g. by Monte-Carlo (MC) techniques such as the PYSHOW algorithm \cite{PYSHOW}, leading to the scale evolution of the FF.  Let us for illustration briefly investigate the physics encoded by PYSHOW. The algorithm treats the QCD shower as an interated series of $1\rightarrow 2$ splitting of a parent parton $a$ into two daughters $b,c$ where the energy is distributed as $E_a = z E_b + (1-z) E_c$ and the relevant virtuality scale (and hence the radiation phase space) decreases with every splitting as parametrized by $t = \log{Q^2}/{\lambda_{QCD}^2}$ until it reaches a non-perturbative virtuality $Q_0 \sim 1$ GeV.

The differential splitting probability at a scale $t$ is then given by the splitting kernel $P_{a \rightarrow bc}(z)$ which can be computed in pQCD for the characteristic subprocess, integrated over the kinematically available range in $z$
\begin{displaymath}
I_{a\rightarrow bc}(t) = \int_{z_-(t)}^{z_+(t)} dz \frac{\alpha_s}{2\pi} P_{a\rightarrow bc}(z).
\end{displaymath}
where the kinematic limits available for $z$ depend on a combination of parent and daughter virtualities $Q_{abc}$ and masses $m_{abc}$ via $M_{abc} = \sqrt{m_{abc}^2 + Q_{abc}^2}$ as
\begin{displaymath}
z_\pm = \frac{1}{2} \left( 1+ \frac{M_b^2 - M_c^2}{M_a^2}\pm \frac{|{\bf p}_a|}{E_a}\frac{\sqrt{(M_a^2-M_b^2-M_c^2)^2 -4M_b^2M_c^2}}{M_a^2} \right).
\end{displaymath}
It is this restriction of radiation phase space due to quark mass which makes the fragmentation in vacuum for heavy quarks much harder than for light quarks.

\section{The QCD shower in the medium}

The uncertainty relation suggests that the formation time of a virtual parton $a$ is parametrically $\tau \sim E_a/Q_a^2$ in the c.m. frame. This allows to argue that the medium can affect neither the hard process itself nor the initial hard branchings in the shower but only later subsequent branchings. Thus, conceptually including the medium corresponds to replacing the FF by a medium-modified FF (MMFF) (see e.g. \cite{ElossPhysics} for a more comprehensive discussion). There are various proposals how the medium might affect the shower, some assume a modification of the splitting kernel leading to fractional energy loss \cite{Q-PYTHIA}, others add to this elastic scattering of partons with the medium \cite{JEWEL}; so far the best tested idea is however to modify the available radiation phase space by introducing a flow of energy and momentum between medium and jet as exemplified by the MC code YaJEM \cite{YaJEM1,YaJEM2}.

Such a perturbative picture of the medium-modified shower leads to an impressive description of multiple observables, in particular it describes correctly the pattern of hadronizing medium-induced radiation double differentially in $P_T$ and angle with respect to the jet axis \cite{jet-h} as observed by the STAR collaboration in jet-h correlations \cite{STAR-jet-h}, the evolution of the nuclear suppression factor from RHIC to LHC \cite{ElossPhysics,RAA} or the full range of energy dependence of the dijet imbalance \cite{A_J} as measured by CMS \cite{CMS}. No viable description of these observables has been suggested so far which is not based on medium perturbations of a perturbative shower picture, and pQCD based jet quenching works across a substantially larger range of observables than discussed here. This striking success of a pQCD picture for light quarks thus suggests that heavy-quark showers (where in addition a hard mass scale exists) should also be tractable by the same techniques.

\section{Heavy-quark showers}

The interactions of QCD are flavour-blind, i.e. gluons couple only to the color charge to quarks, and hence there is no fundamental difference between light and heavy quark jets. Nevertheless, the presence of a substantial quark mass can change the kinematics. As mentioned before, a quark mass introduces a constraint on radiation phase space, thus generically suppressing the vacuum as well as the medium induced QCD radiation (this is known as 'dead cone effect' \cite{DeadCone}) and led to the expectation that heavy quarks should lose less energy due to interactions with the medium than light quarks). Also the formation time of heavy quark showers can be estimated to $\tau \sim E_{jet}/\sqrt{Q_0^2+m_q^2}$, i.e. a $b$-quark shower terminates much earlier than a light quark shower.

Another important aspect is that $m_q \gg T$ with $T$ the medium temperature, i.e. there are no thermally excited heavy-quarks in the medium, which makes conversion reactions like $b\overline{b} \rightarrow gg$ impossible (note however that such conversions are a numerically small effect even for light quarks \cite{Jussi}). Furthermore, heavy quarks are tagged \cite{HQEloss}, i.e. once a hard gluon taking 90\% of the quark's energy is radiated, that energy is lost from the quark, whereas the same emission from a light quark would be interpreted as a change in the identity of the leading light parton since gluon jets cannot easily be identified. Finally, note that for $E\sim$ few GeV light quarks are highly relativistic and can be assumed by move with a velocity $v \approx 1$, where such an assumption can not be made for $b$-quarks.

Since for $Q^2 \gg m_q^2$ the shower evolution is not influenced by the dead cone effect, one can use the uncertainty relation for the shower formation length $L \sim E_{jet}/m_q^2$ to establish conservative estimates for what jet energies $E_{jet}$ a shower surely probes the whole medium ($L>10$ GeV) and when the shower terminates before the medium forms ($L<0.5$ GeV). Using these relations, with $E_c > 78$ GeV and $E_b > 900$ GeV, no differences between the medium modification of heavy and light quark jets are expected as they probe the medium for the same length in a kinematical regime where the quark mass does not matter whereas for $E_c < 3.9$ GeV or $E_b< 45$ GeV one finds that the shower is over before the medium is formed, i.e. in this situation an on-shell quark rather than an evolving shower is the relevant state which is perturbed by the medium. Between these scales is an intermediate region of complicated physics in which one can neither assume that a shower treatment alone nor an on-shell treatment alone would lead to the right physics, and comparison with experiment is neeed to understand which approximation works best. 

Despite claims in the literature that heavy quark energy loss would be a good benchmark test for models designed for light parton high $P_T$ physics, this is hence not true in general but rather depends on the kinematic regime under consideration, as the relevant physics processes can be qualitatively different.

\section{The heavy quark puzzle}

While the dead cone effect discussed above would suggest that the medium is able to induce less radiation (and hence less suppression) for heavy quarks as compared with light quarks, the experimental observation of non-photonic single electron nuclear suppression factor (with the electrons predominantly originating from heavy quark decays) showed a suppression to the same level as for charged hadrons \cite{HQexp1,HQexp2}. This is known as the 'heavy quark puzzle'.

Since the dead cone effect works strongest for the $b$-quark, early attempts at an explanation argued that the ratio of $c$ to $b$ contribution to the electron spectrum isn't well understood, however template fits to the decay kinematics later extracted experimental ratios compatible with theory expectations. Another proposal was that elastic energy loss is quite strong for all partons, i.e. close to 50\% of all energy loss is not via medium-induced radiation but through elastic recoil of medium constituents \cite{HQPuzzle}. However, for the light quark sector, the contribution of such elastic reactions (which would be incoherent) is very well constrained by pathlength dependent observables such as the reaction plane dependence of $R_{AA}$ or the strength of back-to-back dihadron correlations \cite{ElossPhysics,YDE,JussiRP} to be about 10\% of the total.

\begin{figure}
\begin{center}
\epsfig{file=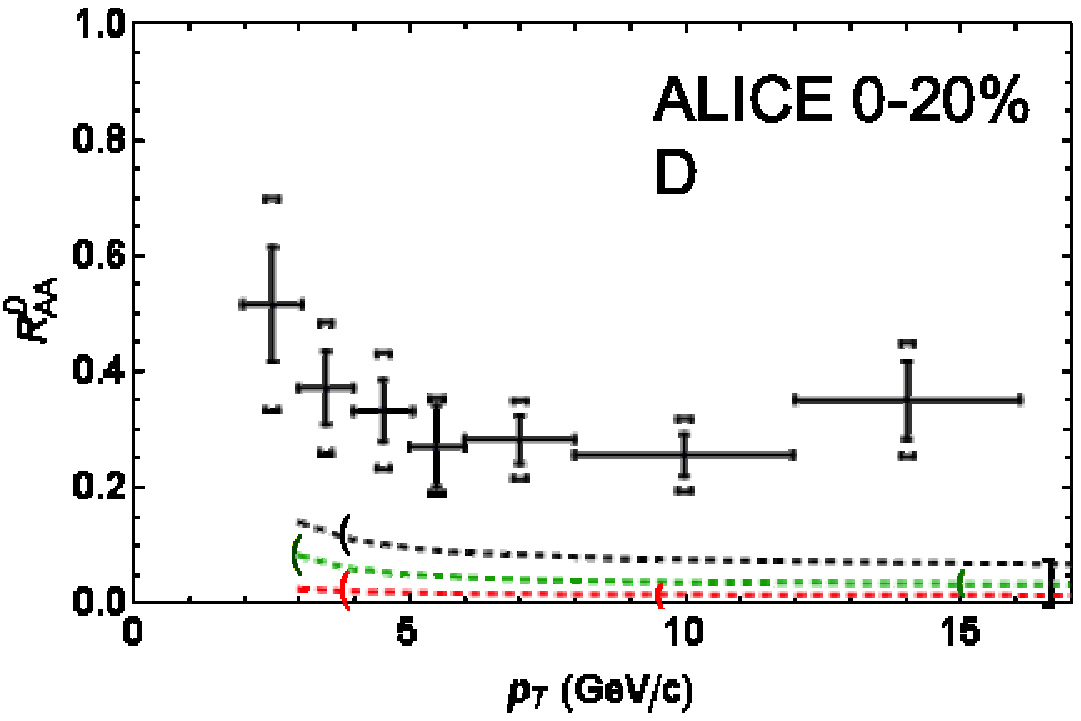, width=6.5cm}\epsfig{file=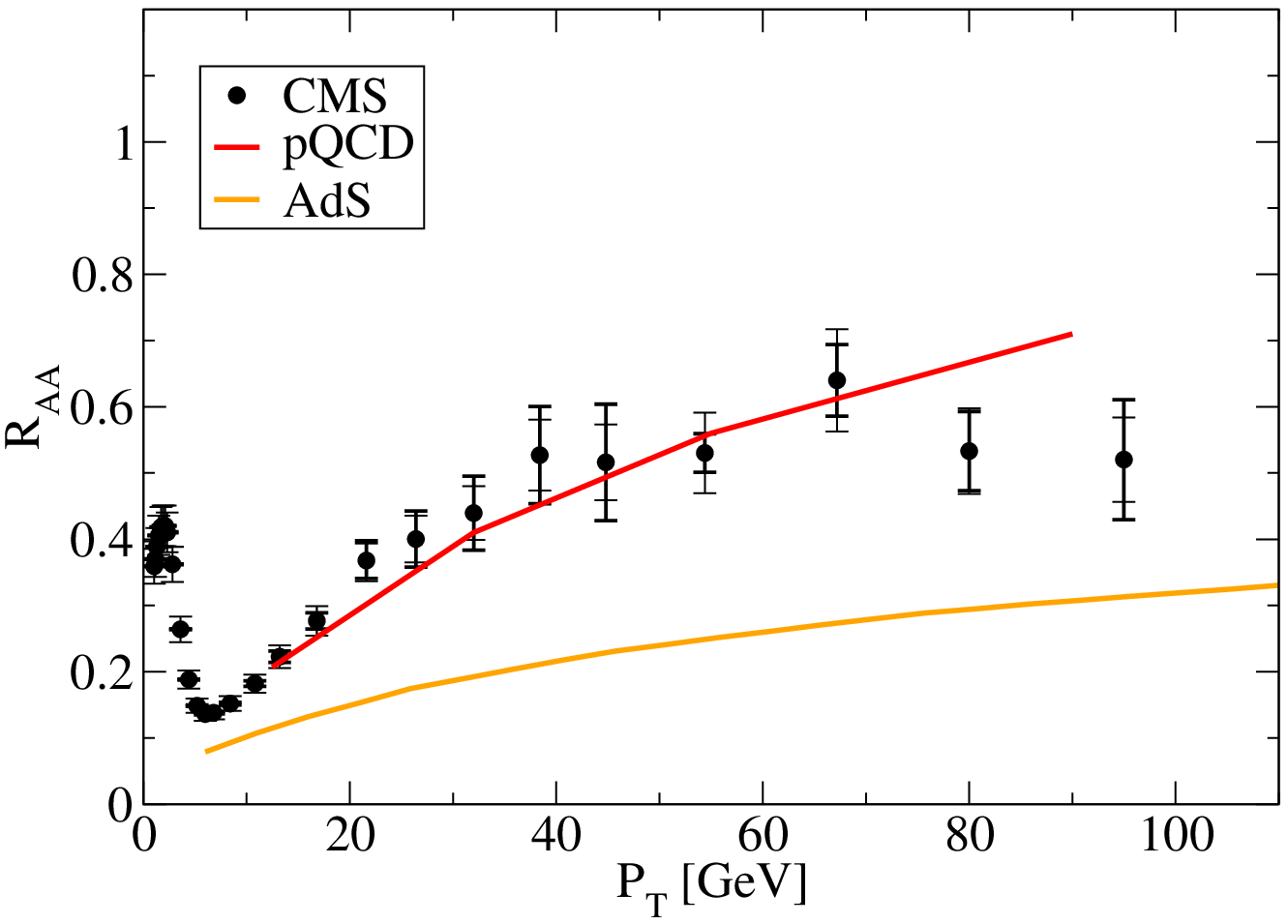, width=6cm}
\caption{\label{F-1} Failure of AdS/CFT strong coupling techniques to extrapolate $R_{AA}$ from RHIC to LHC energies for heavy quarks (left)\cite{HQAdS3} and light quarks (right) \cite{ElossPhysics}}
\end{center}
\end{figure}

A more radical proposals argue that they physics of in-medium showers (especially for heavy quarks) might better be described by strong coupling techniques using the AdS/CFT duality \cite{HQAdS1,HQAdS2}. However, with the first data from LHC available, it could be established that strong coupling techniques fail to predict the evolution of the medium effect from 200 AGeV to 2.76 ATeV correctly, i.e. they overpredict the quenching both for heavy quarks \cite{HQAdS3} and light quarks \cite{ElossPhysics} (cf. Fig.~\ref{F-1}), and to date there is no viable strong coupling model which can account for even a fraction of the large body of high $P_T$ data available.

The currently most promising idea to solve the puzzle involves a selective enhancement of the heavy quark elastic cross section with the medium via resonance formation --- heavy quarks may form proto-hadrons even in a deconfined medium, and while these are unstable, they may still leave a  pronounced effect in the scattering cross section \cite{Resonant1,Resonant2,Resonant3}. Such an effect could work selectively for heavy quarks since light quarks have a long shower formation time and typically do not come on-shell throughout the medium evolution time.

\section{Current heavy-quark phenomenology at high $P_T$}

While modern calculations for the high $P_T$ phenomenology of light partons are mainly concerned with QCD shower evolution and vacuum and medium-induced raditation, heavy quark phenomenology is usually computed with a focus on the elastic reactions of on-shell quarks with a medium. As discussed above, these distinctly different approaches are well motivated by the respective kinematical regimes of the observables.

Usually a $K$-factor is used to enhance the heavy quark elastic cross section in such computations which is then fit to data, aiming for a simultaneous description of the $P_T$ dependence of both the mean electron nuclear suppression factor $R_{AA}$ and its angular dependence with respect to the event plane in terms of the second harmonic coefficient $v_2$. Such approaches may include only elastic reactions (e.g. \cite{Pheno2}) or include medium-induced radiation, albeit suppressed by the dead cone effect \cite{Pheno1} and are able to achieve a fair agreement with the available data. Of particular importance here is the use of realistic and constrained models of the medium evolution. In the light parton sector, it could be demonstrated that the choice of the medium evolution model has a large influence on the obtained value of $v_2$ \cite{HydSys}, and a corresponding study of the systematical effect of the influence of the choice of the medium model on $v_2$ for heavy quark observables is crucial to judge models correctly against the data.

\section{Outlook}

In many computations of elastic energy loss, a fixed value of the QCD coupling $\alpha_s$ is used. However, in reality the running of $\alpha_s$ with the momentum scale is a well established phenomenon, leading to the problem that in the low momentum transfer scatterings expected with medium constituents, $\alpha_s$ may run into a non-perturbative region. One proposal for the determination of the coupling is to obtain it self-consistently with the scale of the Debye screening mass which limits small-angle scatterings in a thermal medium \cite{RunningAlpha} via

\begin{displaymath}
m_D^2 = 4 \pi (1_\frac{1}{6}n_f) \alpha_s(m_D^2) T^2.
\end{displaymath}

Numerically such a procedure results in an enhancement of the elastic scattering cross section as compared to fixed $\alpha_s$ computations. However, the same enhancement would also be relevant for the light quark sector, leading again to a large fraction of elastic energy transfer into the medium which is not observed (note that since a radiation vertex in a shower probes $\alpha_s(Q^2)$ for a high virtuality scale, the same enhancement does not apply to radiative processes). Ultimately, this raises the question of what degrees of freedom in the medium light and heavy quarks are scattering off. Both the success of fluid dynamics in the description of the bulk medium evolution (which can not be obtained in perturbation theory) and the failure of models with a large component of elastic energy loss, assuming scattering off a gas of perturbative quarks and gluons (e.g. \cite{JussiRP}) strongly suggest that the medium degrees of freedom are not light quasiparticles. A precise determination of the energy transfer by elastic reactions both in the light and in the heavy quark sector might provide an important clue as to their true nature.

The current physics picture of light quark phenomenology being described in terms of medium-modified shower evolution vs. heavy quark phenomenology being descibed in terms of largely elastic processes for on-shell quarks which are enhanced by resonant processes is rather compelling for the momentum regimes currently under investigation. The CMS collaboration has openend a new frontier using $b$-tagged jets, demonstrating that at 100 GeV $b$-tagged jets show a $b$-tagged jet $R_{AA}$ compatible with untagged jet $R_{AA}$, a possibly indication that at this scale $b$-quarks might already effectively behave 'light'. 

One crucial test whether the picture of the dominance of elastic energy loss is correct or not are back-to-back correlations which can be shown to be very sensitive to quantum coherence effects and the resulting modifications of pathlength-dependence of the medium modification \cite{ElossPhysics,YDE,ElPhen}. If the observed suppression of non-photonic electrons is indeed the result of elastic scattering processes, then the away side conditional yield ratio $I_{AA}$ of medium over vacuum e.g. $D$ meson yield, given a triggered $D$ meson should be much higher than in the case of, e.g. $\pi-\pi$ correlations. Measuring and successfully describing such correlations would thus go a long way in confirming our current consistent understanding of light and heavy quark phenomenology at high $P_T$ and would open the path for a precision determination of medium properties in the future.

\end{document}